\documentstyle[12pt]{article}
\def\'#1{{\accent19\ifx #1i \i\else #1\fi}}

\def\be{\begin{equation}}
\def\ee{\end{equation}}
\def\bea{\begin{eqnarray}}
\def\eea{\end{eqnarray}}

\newbox\Ancha
\catcode`@=11
\newdimen\ex@
\title{Relationship between the wave function and space}
\author{J. Besprosvany}
\date{Instituto de F\'{\i}sica, Universidad Nacional Aut\'onoma de M\'exico,
Apartado Postal 20-364,  M\'exico 01000, D. F., M\'exico}

\begin{document}
\maketitle
\jot = 1.5ex
\parskip 5pt plus 1pt
\def\baselinestretch{1.9}
\begin{abstract}
We criticize the current standard interpretation of quantum
mechanics, review its paradoxes with attention to non-locality,
and conclude that a reconsideration of it must be made. We
underline the incompatibility of  the conceptions ascribed to
space of {\it field},  and {\it stage} in modern theories, with
differing roles for {\it coordinates.} We hence trace the
non-locality difficulty to the identification of the basis space
of the wave function and physical space. An interpretation of the
wave function in which space loses its {\it stage} use at the
local level, and its physical ({\it field}) meaning is assigned to
the wave function,
 can solve this difficulty. An application of this proposal
implies  a field-equation extension based on a unified description
of bosons and fermions  able to provide new information on the
standard model.

\end{abstract}
 \vskip 1cm
 \baselineskip 22pt \vfil\eject
\section{Introduction}
Quantum mechanics (QM) is a successful theory describing phenomena
in many ranges;    it is also  the  standard framework for the
study of elementary particles, when mixed with relativistic
postulates through relativistic quantum mechanics and  quantum
field theory (QFT). QM's ability of describing the constituents of
nature lends support to its validity at a fundamental level.  In
general, QM faces no challenges on its capacity to describe nature
in principle, except for the on-going and still open question of
how to integrate it with general relativity (GR), and hence to
include gravity in the description. Although the status of the
theory remains solid
  in
its applications at the operational level, paradoxes, confusion,
and doubts linger on its  conceptual basis, its interpretation,
and even its consistency. In particular, these doubts have
remained ever since the so-called Copenhagen interpretation
imposed itself as the standard.
 Postulates of this interpretation
continue to be  questionable and cannot be satisfying for they
lead to the renunciation of objectivity, determinism, and hence,
ultimately to the impossibility of apprehending reality. At the
center of this interpretation lies Bohr's dictum  that  considers
``the space-time coordination and the claim of causality, the
union of which characterizes the classical theories, as
complementary but exclusive features of the
description..."$^{\cite{Bohr}}$. Underlying these notions is
Bohr's stress that   the nature of our perceptions  forces  all
experience  and links to experiment to be formulated in classical
terms, which  evokes  the  Kantian a priori categories. This
interpretation also assumes     the idea of an inherent property
of nature which forbids  assigning physical  meaning to the
variables describing the objects one is measuring (until they are
measured).

 One of the most poignant
debates that followed this interpretation has  centered precisely
on this matter, namely, on whether the wave function carries the
necessary information about an object,  and in general, whether
the quantum mechanical description  can be considered complete.
This question was raised by Einstein, Podolsky, and Rosen
(EPR)$^{\cite{EPR}}$, whose forceful negative answer was argued
with a  gedanken experiment in which they claimed a particle's
variables should be assigned  before a measurement was performed,
while   QM forbids this assignment unless a  measurement is
carried out; this meant either an immediate transmission of
correlations among particles through space, which was discarded as
unphysical, or the conclusion that QM is not complete. Bohr
considered sufficient the explanation  that it is the experimental
set-up which defines the measured physical quantities, and  that
this makes questions about the  particles' state before
measurement meaningless. The EPR argument, on the other hand,
implies  the necessity of hidden variables that an alternative
theory would support. Nevertheless, a later development of which
both Bohr and EPR were unaware was a proof presented by
Bell$^{\cite{Bell}}$
 that the probability predictions of any  local hidden variables theory  should satisfy a  series
  of  inequalities which are violated by QM.
 Moreover, these inequalities can  be subjected
to experimental verification which was actually
performed$^{\cite{AspectaCo}}$, with
 results in accordance with QM.

Thus, the   EPR criticism has motivated  an unexpected
development  in the sense that
 it   has  led to
an additional confirmation of  the  validity of  QM, but with
the implication of the puzzling presence of ``spooky"
correlations in nature,
 that is, an inherent non-local behavior.
In general,  the  Copenhagen  positivistic interpretation
cannot be satisfying  because  it  renounces determinism and
 objectivity, which makes
   the EPR criticism
understandable, but   QM's practical successes preclude yet the
necessity of another theory.

In order to remedy this and other  illnesses a variety  of
alternative interpretations have been constructed, but those based
on the same  questionable  premises  make questionable conclusions
too. The interpretations range from assuming all the consequences
of the Copenhagen interpretation of QM to assuming a purely
classical view. At the first    extreme, for example, the
many-worlds interpretation of QM$^{\cite{Everett}}$  cures the
discontinuity in the collapse of the wave function with the
assumption that, upon measurement,  different states go to other
worlds. However, by underlying  the consistency, this
interpretation  sacrifices veridicality and seems more outlandish
than the nature of the problems in the theory it intends to
correct. At the other extreme
  lies
Bohm's$^{\cite{Bohm}}$ which clings to old classical concepts. For
the purpose of  maintaining an ontological interpretation  of the
objects involved in the quantum mechanical description, it
requires an unlikely understanding of terms as potentials with
unphysical properties as being sourceless fields which violate the
superposition principle. Although these  interpretations can
contain the usual quantum mechanical experiments in their
framework, their  use is less concise than the standard one's, and
they have not provided any new insight into physical problems.

 It is our view that
 to  renounce to scientifically questioning and debating phenomena
 beyond
  what is presently experimentally perceived
cannot be sustained from the QM theory itself and is  therefore
unjustified;
 indeed, the closed and self-contained nature of  the Copenhagen  interpretation has
prevented a discussion on what should be a central theme of
physics.
 Hence,  its
assumption   of a physical  inherent impossibility  of learning
more about an experimental situation
  has represented in a self-prophesying way  only  an impediment for  further research into the matter.
 Doubts emerge on whether the complementarity postulate is a scientific statement, or an unwarranted
 physical and epistemological assumption.
   Actually, both the Bohr  and  EPR views
are based  on  too close a reliance on classical tenets, assumed
to  be the necessary language of natural phenomena,
 and which have been held sacred, as in the case of the concept of  space \footnote
 {This statement should be understood literally
 in the case of Newton.}.
 Yet  the preeminence of quantum phenomena suggests
  classical concepts need not be  the only  way
to describe  experiments. Also, the argument we have presented
above implies the  standard interpretation  of QM is not
satisfying and must be modified and it supports the opinion that
the    accepted  concepts
 of  the wave function and/or space are suspect.

 In this paper we propose a new interpretation of the wave function
   which requires  a modification  of the concept of space used in its
   description, and  in which
 both space and the wave function are assumed to be related.
In Section II we make a  brief  historical review of the concepts
of space and  also an analysis of  its current   conceptions in
modern theories, and in particular,  QM's use of configuration
space and physical  space. In Section III, we ponder some
consequences of the assumption that the latter two are not equal.
In particular, this idea constitutes a possible solution to the
 problem of non-locality in QM.  We consider
also the implications of the proposal  that the wave function and
space are related
 in connection  to the incompatibilities in
modern-theories' conceptions of space.
   The new interpretation  also motivates a new   formulation of field
equations on  an extended spin space,  in Section IV, providing a
unified description of bosons and fermions. In Section V we draw
some conclusions.

\section{A brief history of  space  concepts, and current ones}

A persistent puzzle through the  centuries has been the nature of
space and its relation to physical phenomena.   Controversies have
arisen both around   its form, whether it is finite or infinite,
open or  closed, and  its substance,  whether it is continuous or
discrete, empty or full, and ultimately whether it represents at
all any physical\footnote{That is, related to something one can
measure.}  phenomenon. A closely related debate to the latter
question is the relation of space to the matter that moves in it.
This debate can be summarized into two differing views. In the
``monistic" view space is inseparable and indistinguishable from
the matter that  exists in it, and the distinction between space
and matter is simply a  question of convention. In the
``reductionistic" view space is  of a wholly different nature from
matter, if at all, and is  mainly the receptacle where bodies
exist. While the first view is aesthetically and philosophically
more appealing as it conforms to a unified view of nature, the
second is artificial but  more intuitive, and has been more
successful and useful, by providing a simple framework, or {\it
stage}, to treat phenomena, as in Newtonian   mechanics in
contrast to Cartesian. However, while within the first view space
has a physical meaning doubts remain on whether this is the case
for the second view. For example, Leibnitz argued in this
direction by stating that space is only a system of relations
between bodies.

In the nineteenth century  the debate  centered around  the newly
introduced  field concept, needed to account for extended
phenomena  through space, coming from a novel understanding of
electromagnetism  by Faraday. Formalization
of this concept in Maxwell's equations resulted in an   understanding 
of light as an extended electromagnetic perturbation   through
space.
 Analogy of the behavior  of
light with that of other waves in other media  led physicists to
conclude that space  was a medium,  the ether, an assumption which
supported the monistic view. In a Galilean framework, this medium
would define a preferred  frame of reference for the universe.
Special relativity (SR),
 however,  avoids giving special significance to any particular reference frame through
a new understanding of time\footnote{We shall not discuss the
conceptions of time, although as implied here, they do have an
influence  on the evolution and perception of theories.}, and yet
accounts for electromagnetic phenomena. Thus, it deprives space
again of any relevance except for providing for the {\it stage}
where events occur. This theory  then generalizes the Newtonian
view of space and time into the Minkowskian framework,  but keeps
the basic feature of  using {\it coordinates} to identify the
physical but otherwise  inert points of space and time which
define the bodies' position.

In the twentieth century the debate has continued as the accepted theories of nature subscribe 
to both views. This is the case of  GR, in which   the   local and
global manifestations of space are considered physical. On the one
hand, it is inherent to this theory that, at the local level,
space-time is as a Minkowskian physical framework, or {\it stage},
in which objects fall freely and physical phenomena are the same
as in flat space, independently of the particular spot in which
they occur (second view). On the other hand, at the global level
space-time is a manifold  described by the metric {\it field},
which embodies gravity, and it is coupled to matter. Here {\it
coordinates} are labels to account for particular points in the
manifold but any particular choice lacks physical significance.
(This is expressed by the coordinate invariance of GR). 
The inseparability of space, the gravitational field and matter
gives space the  status of  a physical object
 (first view).
The understanding of space as a {\it field} strongly suggests a
link to electromagnetic phenomena. This possibility was  explored
by Kaluza and Klein who, by extending space to more dimensions,
 constructed  a  model which encompasses both four-dimensional   space-time
and another dimension accounting for the electromagnetic field.

Classical (CM) and quantum mechanics   subscribe to the  second
view. In QM, the wave function, which contains  all the
information of the matter it describes, is defined as a field. The
same is true in QFT, which describes varied numbers of particles
by allowing for an infinite number of degrees of freedom
represented by its principal element, the quantum field. The
latter has space-time as a parameter and satisfies causality
constraints from SR. However,  there is an ambiguity inherent to
the quantum mechanical  treatment with regard to the physicality
of space.  On the one hand, the  wave function's expression in
configuration space represents merely a  basis choice
 and is by no means compulsory; indeed, momentum space
is another possible basis, which means one    is not more relevant
than the other (the incompatibility of these bases leads to
Heisenberg's uncertainty relations). On the other hand, the
association of a particle with a given position comes only after
the wave function is ``collapsed" on that point, that is, when a
measurement is carried out.

\section{Wave function and  space integral view}

 The association of QM's configuration-space basis of the wave function and space, which has been assumed
natural, is by no means obvious or  necessary. Thus, one can
deprive these {\it coordinates}  from their   {\it stage} meaning.
 A notable
implication from  this separation assumption is  a possible
solution to the EPR paradox and Bell's implied non-locality, for
distance loses its universality. A whole set of possibilities for
new conceptions opens up, although these should be  restrained by
the requirement of causality and locality,  whose successful
consequences in QFT do not enjoin their renunciation. One possible
path to follow is that if
 configuration space is deprived of a direct
association with physical space, having    only an ascribed
meaning of
 basis {\it coordinate}, the remaining physical quantity
 left in the quantum mechanical description,
the wave function,  must contain the complete  information on both
matter and space. We call integral  this view of space  and the
wave function.

Generally,  under the  conception  of space as   {\it field}, as
in
 global  GR, and in  QM and QFT--which deny any physical meaning
to a well-defined place where objects are--{\it coordinates} play
merely a
 descriptive role. But under the
conception of  space as {\it stage}, as   background of events in
the local description of GR, and in CM, QM, and QFT (after a
measurement is performed), {\it coordinates} are ascribed a
physical meaning.

Thus, the above proposal overcomes  the incommesurable uses of
{\it coordinates}  in the   {\it field}  and  {\it stage}
descriptions of space, underlines  their use of
  as a  pure bookkeeping device, and fits  a preferable unified  view of
space as field   and the idea that space is not void but is a
manifestation of a ``space-matter" substance.

 With the identification of  the wave function with space,   the latter
  acquires a   {\it field} meaning, described by
 {\it coordinates},  whose  {\it stage} meaning is
dropped. Hence, by subscribing to the view that the wave
function, which  describes  matter,    fills up space  we dispose
of space as {\it stage} at the local level too. In this way, we
are emulating the treatment of global GR  by interpreting  the
wave function (space) as the relevant field, whose coordinates do
not have a direct physical meaning; at the same time we apply
this idea at the local level, which  GR does not do. This idea therefore
brings  QM and GR nearer and allows for a removal of an
inconsistency in an entirely new framework which gives space a
new meaning locally. However, the analogy between GR's space and
QM's configuration space is only partial, given that the
multiplicity of particles requires in principle a multiplicity of
configuration coordinates. Also, while a coordinate
 in GR describes a given  space-time point with a given metric property, in the proposed
QM interpretation  all physical meaning is assigned to the wave
function and coordinates become labels with the possibility of
redundancy in the description.  We may expect that in the
classical limit space regains its usual meaning of {\it stage}.
The connection to this description  should come through  average
quantities such as the two-particle density-matrix.

\section{Boson and fermion field equations on an extended spin space}

A shared description is expected in a  unified treatment of the
wave function and space. Hence, it is plausible to have
 fermions and space, whose representing field, the
graviton, is a boson,  under the same footing. This suggests a
closer connection between fermions and bosons, and in general,
matter fields under the same footing as interaction fields. To
implement this idea we need  a formalism that relates a field to
the very structure of space-time: this link is present   at a
fundamental level for fermions    through Dirac's equation and its
related matrices, which use  the simplest $SO(3,1)$
representation. Through these matrices we expect a link between
the symmetry of space-time, and the fermions that the equation
describes,
 that is, a
link between the structure of space and matter. To the extent
that this description can include bosons we hope some information
will be obtained about gauge interactions and their vector
particles, and eventually, about spin-two particles, the carriers
of gravitation. Indeed, while the  ultimate goal of a unified
theory may be   to construct it  as encompassing  that the
curvature of space in  GR be linked to the wave function,
 at present  we  concentrate on a more modest   Lorentz-invariant
approach using  Minkowski's space-time. Thus, we search for a
description of vectors and scalars as close as exists for
fermions in order to be able to relate both representations. We
also demand that the  field equation which provides such
description be  enclosed in a variational principle framework.
These requirements   are attained by generalizing Dirac's
equation and by extending its multiplet content.
 Then,
 instead of assuming the Dirac operator acts on a
 spinor$^{\cite{Dirac}}$
\begin{eqnarray}
\label {Diraceq} ( i \partial_\mu\gamma^\mu -M) \psi=0,
\end{eqnarray}
where $\psi$ is the column vector with components $  \psi_\alpha$,
 we  assume  it
acts on a $4\times 4$ matrix $\Psi$ with components $\Psi_{\alpha
\beta}$, so that the equation becomes
\begin{eqnarray}
\label {Jaimeq} ( i \partial_\mu\gamma^\mu -M)\Psi ={ 0}.
\end{eqnarray}There are, then, additional possible transformations and symmetry operations
 that further classify $\Psi$.
 The Dirac-operator transformation $( i \partial_\mu\gamma^\mu -M)\rightarrow
{U}(\it i
\partial_\mu\gamma^\mu -M) {U^{-1}}$
 induces
the left-hand side of the transformation
\begin{eqnarray}
\label  {transfo}
 \Psi\rightarrow U \Psi U^\dagger,
\end{eqnarray}
and $\Psi$ is  postulated to transform  as indicated on the
right-hand side. Thus, all symmetry operators valid for the Dirac
equation in eq. \ref{Diraceq} (with its corresponding particular
cases of massless and massive cases)
 will be valid as well for it. The operators therefore satisfy
the Poincar\'e algebra.

 That the equation, the transformation and symmetry operators $U$,
 and the solutions $\Psi$  occupy the same space is not only
 economical
but  it befittingly   implements  quantum mechanics, for it
ultimately implies measuring apparatuses are not constituted
differently in principle from the objects they measure.

  $U$ and $\Psi$ can be classified in terms of  Clifford
algebras. In four dimensions (4-$d$) $U$ is conventionally a
$4\times 4$ matrix
  containing symmetry operators as the Poincar\'e generators,
  but it  can contain others, although,     e. g., in the  chiral  massless case
it   can only carry an additional $U(2)$ scalar
symmetry$^{\cite{Jaime}}$.
More symmetry operators appear
 if Eq. \ref{Jaimeq},  $\mu=0,...,3,$ is assumed within  the larger Clifford
algebra ${\mathcal C}_{N},$
 $\{ \gamma_\mu ,\gamma_\nu \} =2 g_{\mu\nu}$, $\mu,\nu=0,...,N-1$, where   $N$ is the
 (assumed even) dimension,
 whose structure is
  helpful in classifying the available symmetries, and which
  is represented  by  $2^{N/2}\times 2^{N/2}$
 matrices.
  The  usual 4-$d$ Lorentz symmetry,  generated in terms of
   $\sigma_{\mu\nu}=\frac{i}{2}[ \gamma_\mu,\gamma_\nu
 ],$  $\mu ,\nu =0,...,3$, is
 maintained and
  $U$  contains also $ \gamma_a$, $a=4,...,N-1$,  and their
 products as possible symmetry
 generators.
Indeed, these elements are  scalars for they commute
 with the Poincar\'e generators, which contain  $\sigma_{\mu\nu}$, and
they are also  symmetry operators  of  the massless  Eq.
\ref{Jaimeq},  bilinear in the $\gamma_\mu$ matrices, which is not
  necessarily the case for mass terms (containing  $\gamma_0$).
 In addition, their products with
   $\gamma_5=-i\gamma_0\gamma_1\gamma_2\gamma_3$  are Lorentz  pseudoscalars.
 As
 $[\gamma_5,\gamma_a]=0$,
  we can classify the (unitary) symmetry algebra as
  ${\mathcal S}_{N-4}={\mathcal S}_{(N-4)R} \times {\mathcal S}_{(N-4)L},$
  consisting of the projected
 right-handed ${\mathcal S}_{(N-4)R}=
\frac{1}{2}(1+\gamma_5)U(2^{(N-4)/2})$ and left-handed ${\mathcal
S}_{(N-4)L}= \frac{1}{2}(1-\gamma_5)U(2^{(N-4)/2})$
  components.

The solutions of Eq. \ref{Jaimeq} do not span all the matrix
 complex space, but this is achieved by considering also solutions of
\begin{eqnarray}
\label {Jaimeqnext} \Psi\gamma_0( -i \stackrel{\leftarrow}{
\partial_\mu}\gamma^\mu -M) ={ 0},
\end{eqnarray}
 consistent with the transformation in Eq. \ref{transfo},
 (the Dirac operator  transforming accordingly).

It is not possible to find always solutions that simultaneously
satisfy equations of the type   \ref{Jaimeq} and \ref{Jaimeqnext}
(except trivially), which means they are not simultaneously
on-shell, but they satisfy at least one and therefore the
Klein-Gordon
 equation. Indeed, the solutions of eqs. \ref{Jaimeq} and \ref{Jaimeqnext}, can
 be generally characterized as  bosonic  since
  $\Psi$ can be
understood to be formed of spinors as $\sum_{i,j} a_{ij} |w_i
\rangle\langle w_j |$.

Generalized operators acting on this  tensor-product space (spinor
$\times$ spinor $ \times$ configuration or momentum space) further
characterize the solutions. Positive-energy solutions, according
to Eq. \ref{Jaimeq} are interpreted as negative-energy solutions
from the right-hand side. This problem is overcome if we assume
the hole interpretation for the $\langle w_j |$ components, which
amounts to the requirement that operators generally acting from
the right-hand side acquire a minus, and that   the commutator be
used for operator evaluation.
 Thus, the 4-$d$ plane-wave solution
combination $\frac{1}{4}[(1- \gamma_5)\gamma_0(\gamma_1-i
\gamma_2)]e^{-i kx}$, with $k^\mu=(k,0,0,k),$ is a massless
vector$-$axial ($V-A$) state propagating along $\hat {\bf z}$ with
left-handed circular polarization, normalized covariantly
according to $\langle    \Psi_A
  | { \Psi_B}\rangle =tr\Psi_A^\dagger \Psi_B,$ the generalized inner
product for the solution space. In fact, combinations of solutions
of Eqs. \ref{Jaimeq} and \ref{Jaimeqnext} can be formed with a
well-defined
  Lorentz
index:   vector  $\gamma_0\gamma_\mu$, pseudo-vector
$\gamma_5\gamma_0\gamma_\mu$, scalar $\gamma_0$, pseudoscalar
$\gamma_0\gamma_5$, and antisymmetric tensor
$\gamma_0[\gamma_\mu,\gamma_\nu]$.
  For example,
 ${ A}^{\mathcal C}_\mu(x)
=\frac{i}{2}\gamma_0\gamma_\mu e^{-i    kx}$
 is a combination  that
 transforms  under parity into $ { A}^{{\mathcal C}\mu} (\tilde x ),$ $\tilde x_\mu=x^\mu,$ that is, as a vector.
 We may also view $\frac{1}{2}\gamma_0\gamma_\mu$ as  an orthonormal polarization
 basis, $A_\mu=tr\frac{1}{2}\gamma_\mu A^\nu\frac{1}{2}\gamma_\nu$\footnote{As for
 $\bar \psi=\psi^\dagger\gamma_0$, a unitary transformation can be
 applied to
 the fields and operators to convert them to a covariant form.}; just as
   $n_\mu$ in
  $A_\mu=g_{\mu\nu}A^\nu =n_\mu\cdot A^\nu n_\nu$.
In fact, the sum of  Eqs.  of \ref{Jaimeq} and
 \ref{Jaimeqnext} implies $^{\cite{Bargmann}}$  for a $\Psi$ containing $\gamma_0\slash\!\!\!\! A={ A}^\mu\gamma_0\gamma_\mu $  that  ${
 A}^\mu$ satisfies the free Maxwell's equations.

Solutions   contain also products of $\gamma_a$ matrices
 that define their scalar-group representation.
 For given  $N$, there are variations of  the symmetry algebra
depending on the chosen Poincar\'e generators and Dirac equation,
respectively, through the  projection operators ${\mathcal P}_P,\
{\mathcal P}_D\in {\mathcal S}_{N-4}$, $[{\mathcal P}_P,{\mathcal
P}_D]=0$. ${\mathcal P}_P$ acts as in, e.g., ${\mathcal
P}_P\sigma_{\mu\nu}$, and
 ${\mathcal P}_D$ modifies  Eqs.
\ref{Jaimeq} and \ref{Jaimeqnext}  through ${\mathcal
P}_D\gamma_0( i
\partial_\mu\gamma^\mu -M).$ Together, they
characterize   the  Lorentz and scalar-group solution
representations.   We require     ${\rm rank}{\mathcal P}_D \le
  {\rm rank} {\mathcal P}_P$, for otherwise pieces of the solution
space exist that do not transform properly. For ${\mathcal P
}_D\neq 1$  Lorentz operators  act trivially on one side  of the
solutions containing $ 1-{\mathcal P }_P $,  since $(1-{\mathcal P
}_P){\mathcal P }_P=0$,  so we achieve the goal of having fermions
in in a common description with bosons.

An interactive field theory can be constructed in terms of the
above  degrees of freedom.
We consider a  vector  and fermion non-abelian gauge-invariant
theory. The expression for the kinetic component of the Lagrangian
density  \begin{eqnarray} \label {Lagra}{ \mathcal L}_{V}=
-\frac{1}{4}F_{\mu\lambda}^ag^{\lambda\eta}\delta_{ab} F^{b\mu}_{\
\  \eta}=-\frac{1}{4 N_o}tr {\mathcal   P}_D
F_{\mu\lambda}^a\gamma_0\gamma^{\lambda} G_a F^{b\mu}_{\ \
\eta}\gamma_0\gamma_ \eta G_b\end{eqnarray}
 shows ${\mathcal   L}_{V}$
is equivalent to a trace over combinations over normalized
components $\frac{1}{\sqrt{ N_o}}\gamma_0\gamma_\mu G_a $ with
coefficients $F^a_{\mu\nu}=\partial_\mu A_\nu^a-
\partial_\nu^a A_\mu^a+ g A_\mu^b A_\nu^c C^a_{bc}$, $g$ the
coupling constant,  $\gamma_\mu\in{\mathcal C}_{N}$, $G_a\in
{\mathcal S}_{N-4} $ the group generators, $C^a_{bc}$ the
structure constants,
  and $N_o=trG_a G_a $, where for
non-abelian irreducible representations we use $trG_i G_j=2
\delta_{ij}$.

  Similarly,   the interactive part of the fermion gauge-invariant Lagrangian
$ {\mathcal   L}_{f}=\frac{1}{2}{\psi^\alpha}^\dagger\gamma_0(
i\stackrel{\leftrightarrow}\partial_\mu-g A^a_\mu G_a
)\gamma^\mu\psi^\alpha,$
 with $\psi^\alpha$  a  massless spinor with flavor $\alpha$,
can be written
   ${\mathcal   L}_{int} = -g\frac{1}{ 2 N_o} tr
{\mathcal P}_D A_{\mu}^a \gamma_0\gamma^\mu G_a
j^{b\alpha}_\lambda \gamma_0\gamma_\lambda G_b,$    with  $
j_\mu^{a\alpha}=tr{\Psi^\alpha}^\dagger\gamma_0\gamma_\mu G_a
\Psi^\alpha$
 containing $\Psi^\alpha=\psi^\alpha\langle \alpha |$, and $\langle \alpha|$
 is a row  state accounting for the flavor.
 ${\mathcal   L}_{int}$ is written in terms of  $\gamma_0\slash\!\!\!\! A$,
 and $\gamma_0\slash\!\!\!j^{a\alpha}$,
 that is, the vector field and the current occupy   the same spin space. This
connection and the QFT understanding of this vertex as the
transition operator between fermion states, exerted by a vector
particle, with the coupling constant as a measure of the
 transition probability, produces information on the coupling
 constant$^{\cite{Jaime,Besprosub}}$.


As for the initial formulation, ${\mathcal P}_D$ restricts the
possible gauge symmetries that can be constructed in the
Lagrangian, for $ \gamma_0\gamma_\mu G_i$ needs to be contained in
 the space it projects. Thus,
${\mathcal P}_P$ and ${\mathcal P}_D$ determine the symmetries,
   which are global, and  in turn,
determine the  allowed gauge interactions. Furthermore, they
 fix the representations, assumed physical.
The $N=6$ case has been
 researched$^{\cite{Jaime}}$ and connections have been found to the
 $SU(2)_L\times U(1)$, electroweak sector of the SM.
 as a result,
   restrictions are provided on the representation choices,
 vertices, and coupling constants.  Relevant
information on the standard-model representations and interactions
is obtained from the 10-$d$ case$^{\cite{Besprosub}}$.

Thus, the  extended Dirac equation$^{\cite {Jaime}}$ in Eq.
\ref{Jaimeq}, the Bargmann-Wigner equation$^{\cite{Bargmann}}$,
and  the expression for a standard Lagrangian  as in Eq.
\ref{Lagra}  have in common that fields are formulated on an
extended spin space, with the possibility of relating some
generators in this space to scalar degrees of freedom. This is a
limited but significant example of a consistent generalization
that connects the  space-time and scalar  symmetries, giving a
unified description of boson and fermion fields. It suggests
 a research direction for an ultimate formalism describing space
and the particles' wave function in a unified way.

\section{Conclusions}

In this paper we have proposed  an interpretation   of  QM in
which its  non-local correlation paradoxical aspect, implied by
Bell's inequalities, is removed. This entitles a separation  of
the concept of space as basis, used in the description of the wave
function,  and physical space; then,  only the wave function
assumes a physical meaning and it  encompasses both space and
matter. Our proposal is both radical and conservative for it
advocates a modification of the notion of space which has been
assumed untouchable,  and yet, it  has the aim of satisfying
locality, thus  explaining  instantaneous correlations.
 The idea presented goes beyond being only a conceptual interpretation   for it motivates a formulation  of
 field
 equations  that
has relevant consequences in particle physics and  embodies this
interpretation well.
 We do not claim that within  this interpretation  all QM
paradoxical  aspects will go away for clearly this requires
dealing with the problem of the collapse of the wave function,
which is presently under intense experimental and theoretical
research. Rather, we advocate another conceptual framework
in which problems as  the collapse of the
wave function and randomness can be  confronted afresh.

  The  approach thus presented also generally implies that in
dealing with QM's
 old problems   a researcher armed with mathematical
tools, his imagination, and a disposition to question classical
tenets  could rehabilitate investigations whose aim of picturing
what is going on in quantum phenomena has  not been  considered
productive. Thus, for example, we speculate that within such an
approach the fact that the system is inescapably perturbed would
be just a natural  consequence, and not something that would
impede our capacity for forming a picture of  events.  Also,
Heisenberg's uncertainty relations would be interpreted as not
limiting  our knowledge of reality, as commonly understood, but
only our expectations about how this knowledge should be.

The pervasiveness and physical nature of the wave function have
been proven  in innumerable cases. The arguments presented in this
paper  imply a negative answer to   the  question ``Is the wave
function different from space?" constitutes a viable
interpretation of quantum mechanics that solves some of its
puzzles. This interpretation allows for its simplicity to be kept
by using the framework of space-time coordinates, while these are
stripped of any direct physical meaning.

 The
scientific quest for a unified description of nature is as
ancient as the early Greek philosophers who conceived the
concept  of
 apeiron, a primordial matter of which  objects are  constituted.
 The
interpretation presented here brings closer a description of the
wave function and the space-time it is supposed to traverse.
 Having the fields stemming from a single coordinate base
  brings us closer to the idea that  the (matter
and carrier) fields  are but  aspects of a single entity.

\noindent{\bf Acknowledgements}

The author acknowledges support from  DGAPA-UNAM through project
IN118600 and and CONACYT through project 3275-PE.

\setcounter{equation}{0}


\end{document}